\documentclass[preprint1]{aastex}

\usepackage{natbib}
\def\kms{$\rm{km\,s^{-1}}$}
\def\arcsec{$^{\prime\prime}$}

\def\degree{$^{\rm o}$}
\def\CaII7291{Ca {\sc II}] $\lambda\lambda$ 7291,7323\ }

\def\OI6300{[O {\sc I}] $\lambda\lambda$ 6300,6364\ }

\def\Msun{M$_{\rm \odot}$}
\def\qisp{Q$_{ISP}$}
\def\uisp{U$_{ISP}$}

\def\dm15{$\Delta M_{15}$}
\def\degree{$^{\rm o}$}
\newcommand\gr{$\gamma$-ray}
\newcommand\grs{$\gamma$-rays}
\newcommand\grb{$\gamma$-ray burst}
\newcommand\grbs{$\gamma$-ray bursts}

\def \gta {\mathrel{\vcenter
     {\hbox{$>$}\nointerlineskip\hbox{$\sim$}}}}
 \def\ni{$^{56}{\rm Ni}\ $}      %Ni56 symbol
\begin{document}

\title {
Spectropolarimetry of the Type Ic SN 2002ap in M74: More
Evidence for Asymmetric Core Collapse
\footnote{Based on observations collected at the
European Southern Observatory, Chile (ESO Progr. No. 68.D-0571(A).}
}

\author{Lifan Wang$^{1}$,
Dietrich Baade$^{2}$,
Peter H\"oflich$^{3}$, J. Craig Wheeler$^3$,
Claes Fransson$^4$, and Peter Lundqvist$^4$
}

\affil{$^1$Lawrence Berkeley National Laboratory 50-232\\
    1 Cyclotron Rd, CA 94720}

\affil{$^2$European Southern Observatory\\
    Karl-Schwarzschild-Strasse 2\\
     D-85748 Garching, Germany}
\affil{$^3$Department of Astronomy and McDonald Observatory\\
          The University of Texas at Austin\\
          Austin,~TX~78712}
\affil{$^8$ Stockholm Observatory, AlbaNova, Department of Astronomy\\
SE-106 91 Stockholm, Sweden}

\begin{abstract}
High-quality spectropolarimetric data (range 417-860 nm; spectral resolution 1.27
nm and 0.265 nm/pixel) of SN 2002ap were obtained with the ESO Very
Large Telescope Melipal (+ FORS1) at 3 epochs that correspond to
-6, -2, and +1 days for a V maximum of 9 Feb 2002.  
The polarization spectra show three distinct
broad ($\sim$ 100 nm) features at $\sim$ 400, 550, and 750 nm that
evolve in shape, amplitude and orientation in the Q-U plane.  
The continuum polarization grows from nearly zero to $\sim$ 0.2 percent.  
The 750 nm feature is polarized at a level $\gta$ 1 percent. 
We identify the 550 and 750 nm features as Na I D  
and OI $\lambda$ 777.4 moving at about 20,000 \kms.  
The blue feature may be Fe II.

We interpret the polarization evolution in terms of the impact of 
a bipolar flow from the core that is stopped within the outer envelope of 
a carbon/oxygen core.  Although the symmetry axis remains fixed,
as the photosphere retreats by different amounts in different 
directions due to the asymmetric velocity flow and density distribution 
geometrical blocking effects in deeper, Ca-rich layers 
can lead to a different dominant axis in the Q-U plane.  

We conclude that the features that characterize SN~2002ap, specifically
its high velocity, can be accounted for in an asymmetric model with
a larger ejecta mass than SN~1994I such that the photosphere remains
longer in higher velocity material.  The characteristics of 
``hypernovae" may be the result of orientation effects in a mildly 
inhomogeneous set of progenitors, rather than requiring an excessive 
total energy or luminosity.  In the analysis of asymmetric events
with spherically symmetric models, it is probably advisable to
refer to ``isotropic equivalent" energy, luminosity, ejected mass,
and nickel mass.

\end{abstract}


\section{Introduction}

From the first identification of Type~Ic supernovae (SN~Ic) as a 
distinguishable spectral subclass \citep{WheelLev85} it has been recognized 
that they represent the collapse of bare non-degenerate, carbon/oxygen cores 
\citep{vanDyk:1992,CloccWheel97}.  Most display a range in peak
absolute magnitudes from -16.5 to -18.5, but some may be
considerably brighter, exceeding -19 \citep{Clocc:2000} without
showing excessive velocities or light curve anomalies.  They also 
show a variety of rates of decay from maximum \citep{Clocc83N:1996,
Clocc83V:1997, CloccWheel97, CloccIcLC97}.  
The best guess is that they represent
the collapse of the core of a massive star that has lost most
or all of both its hydrogen and helium layers.  The remaining
core mass and hence ejecta mass is expected to show some dispersion
that will be reflected in the properties of the observed explosions.
In addition, the suspicion that all core collapse supernovae, and even more
so all SN~Ic, are strongly asymmetric \citep{Wang:2001} suggests
that there will be significant observer line-of-sight effects
that can be manifest in the luminosity and photospheric velocity
as well as the spectropolarimetry.  The observed properties, especially
the photospheric velocity, are also expected to be strong functions
of epoch with higher velocities at earlier times.

The advent of SN~1998bw and its apparent connection with GRB~980425
\citep{Galama98} drew new attention to the general class of SN~Ic.
SN~1998bw displayed large photospheric velocities early on and
was a strong source of radio radiation implying some relativistic
ejecta \citep{Kulkarni98}, although an especially weak \grb.
At about the same time \citet{Pacz98} coined the term ``hypernova"
to specifically mean the events with large optical luminosity
that were associated with \grbs\ and their afterglows.  This
term was adopted by some in the supernova community to represent
supernovae, SN~1998bw in particular, that seemed to be excessively
bright and to require very large amounts of kinetic energy and 
ejected \ni\ mass \citep{Iwamoto98bw, Woosley98bw}.  Since that time, the
term ``hypernova" has been used to apply to events that resemble
SN~Ic, but have large photospheric velocities, whether or not
they are especially bright \citep{Iwamoto:97ef, Mazef00}. 
 
These developments raise the issue of whether ``hypernovae" as
the term is applied to hydrogen-deficient supernovae is really
a separate class with excessively large kinetic energy and
ejected \ni\ mass or whether these supernovae can be explained in terms of
the normal range of properties, including asymmetries, associated
with bare carbon/oxygen core collapse \citep{HWW:1999}.
  
SN~2002ap has attracted much attention because early
spectra showed a lack of hydrogen and helium characteristic of
SN~Ic and broad velocity components \citep{KinugasaIAU,
MeikleIAU,Gal-YamIAU}, which, as mentioned above, have been 
taken as one characteristic of ``hypernovae." The nature,
existence of, and import of ``hypernovae" remains to be clarified,
and the study of SN~2002ap presents an important opportunity.
This supernovae was not associated with a \grb, was not
especially bright \citep{Gal-Yamap02, Mazap02}, and was a weak
radio source \citep{Berger:2002}. 
Since its only special property seems to be
a high photospheric velocity at early times, study of the velocity
in the context of geometrical asymmetries may shed light on the
general category of ``hypernovae."

A particularly important issue is to understand whether asymmetries
could be affecting the interpretation of the kinematics, luminosity,
kinetic energy, nickel mass, ejected mass, and progenitor mass
of the supernova.  The prime tool to investigate the geometry of
distant supernovae is spectropolarimetry \citep{Wang:1996,Wang:2001,
Leonard:2000,Leonard:2001a,Leonard:2001b,Howell:2001}.  Here
we present spectropolarimetry of SN~2002ap from about six days before to
about one day after peak visual magnitude.  This is the first
time that the temporal evolution of the polarization has been
studied in a Type Ib/c and the first time that such early
polarimetry has been obtained of a Type Ib/c.

\citet{Wang:1996} studied supernova polarimetry
published before 1996 and found that Type Ia supernovae are normally
not polarized and on average show much lower polarization than
core-collapse supernovae (Type II, Ib/c; but see \citealt{Howell:2001}).
Subsequent data indicated a higher
degree of polarization for the bare-core or low mass events
such as Type IIb, and Ib/c supernovae, than for more massive
ejecta \citep{Wang:2001}. The degree of polarization is
a function of time after explosion.
For core-collapse events, the degree of polarization is higher past
optical maximum than before optical maximum
\citep{Wang:2001,Leonard:2001a,Leonard:2001b}.
\citet{Wang:2001} reported the polarization of the Type Ic SN~1997X
that was exceptionally high, around 7 percent. Some of this is
probably interstellar in origin, but the suggestion remains that
``bare core" supernovae like SN Ic are very highly distorted.  There
was only one epoch of data on SN~1997X.  SN~2002ap gives the opportunity
to study the polarization evolution of a Type Ib/c event in
unprecedented detail.

We present in \S 2 observations and data reduction of
SN 2002ap.  We discuss the spectroscopic and spectropolarimetric evolution
in \S 3. The structure of the SN 2002ap ejecta is studied in \S 4.  
In \S 5, we give some discussion and conclusions with 
emphasis on implications for models of SN Ib/c and ``hypernovae."

\section{Observations and Data Reduction}

SN~2002ap was discovered in M74 on 29 Jan 2002 by Y. Hirose 
\citep{Nakano02IAU}.
It was first detected about 10 days before V maximum.
At a distance of order 10 Mpc, it peaked at U=13.3 mag around 5 Feb,
at B=13.1 mag around 7 Feb, at V=12.4 mag around 9 Feb,
%[Note Mazzali et al say 8 Feb], 
at R=12.2 around 10 Feb, and at I=12.3
around 15 Feb \citep{Gal-Yamap02}
%[note these guys are making their photometry publicly available
%and we might want to use it].
%[note for future reference that Gal-Yam et al take note of
%a transient narrow absorption that they attribute to He II 4876]
The host galaxy, M74 (NGC 628), is a face-on spiral galaxy.  
The supernovae is 258" west and 108" south of the center of M74 
at  R.A. = 1h36m23s.85, Decl. = +15o45'13".2 (equinox 2000.0)
\citep{Kushida02IAU}.  The   
supernova is about 5 arcminutes from the nucleus where the
background contamination from the galaxy and extinction within
the galaxy is likely to be quite low 
\citep{Smartt02IAU}.
Galactic reddening on the line of sight to M74 is also low,
E(B-V) = 0.008 mag  \citep{Schlegel:1998}.
%[NOTE Kawabata et al. give 0.09, I took this number from the
%A-V given by Gal-Yam divided by 3.1]

The observations were obtained in service mode through our Target
of Opportunity program at the ESO-VLT.   We used
Melipal (= UT3) with FORS1 \citep{Wangel:2002}.
FORS1 employs a TK2048EB4-1 2048$\times$2048 backside-illuminated
thinned CCD. The cooling of the CCD is performed by a standard
ESO bath cryostat. This system uses a FIERA controller for CCD
readout. The grism GRIS-300V (ESO number 10) was used for all
observations.
The order separation filter GG 435+31 (4)  was used in some observations
to study the Ca II IR triplet, but was also taken out to observe
spectral features in the blue end.
%LIFAN CHECK THIS
The supernova was always observed in four subsequent separate
exposures with the position angle of the Wollaston prism at 0\degree.0,
45\degree.0, 22\degree.5, and 67\degree.5.
The $\lambda/2$ retarder plates are of the ``superachromatic'' type, and the
position angles can be set with an accuracy of 0.1 degree. The chromatic
zero angles were set by the data provided in the VLT FORS1 user manual
\citep{FORS1}. The spectral resolution for these observations
is around 12.7 \AA\ as measured from the spectral calibration lamps --
with each pixel corresponding to  about 2.6 \AA\ in wavelength scale.
A slit of 1.0\arcsec\ wide was used for  all these observations.

We have also obtained images for bias correction,
wavelength calibration, and  flat fielding for each observing run. The
wavelength calibration and flat fielding  images are all obtained with the
waveplate rotated to the same position angles used  for the supernova
observations, i. e., 0\degree.0, 45\degree.0, 22\degree.5,
67\degree.5.
The flat field images of each run are combined to form the final flat field
correction image that is applied to each data set.
Standard data reduction procedures including bias correction, flat-fielding,
background subtraction, spectrum extraction, wavelength calibration, and
flux calibration were performed using the IRAF package \citep{Tody93}.
The polarized
spectra were reduced using our own software. The per-pixel photon
statistical errors of the Q and U vectors of the VLT data were typically
below 0.1\%. Stokes parameters were
rebinned to a 15 \AA\ interval \citep{Wangel:2002}. 
The binning was done by calculating
the average polarization within a bin weighted by the number of detected
photons within each pixel \citep{WWH:1997}.
The 15 \AA\ binning was chosen so that it is slightly larger than 
the spectral resolution of the data (12.7 \AA) 
to eliminate sampling errors of the resolution element. The rebinning is
equivalent to calculating the degree of polarization of all photons
integrated in each bin; when the bin size is larger than the
resolution element, the errors of the Stokes Parameters in
each wavelength bin are uncorrelated.  
Typical photon statistical errors in the Q and U
vector after binning to the 15 \AA\ interval are around 0.02\% for all of
these observations and are not a major source of error in the discussion of
these data. During the commissioning of FORS1 in 1998/1999 the systematic
instrumental polarization of FORS1 was found to be less than 0.1\%
(Szeifert 1999, private communication). Errors during data reduction are
expected to be small as well ($<0.02\%$) as the data are reduced in double
precision.

\section{Spectropolarimetric Properties of SN 2002ap}

Optical spectra of SN~2002ap have been presented by \citet{Gal-Yamap02} 
at days -9, -2, +5, +12, and +18 with respect
to V maximum, and an optical spectrum obtained by Meikle at +1 d is 
presented by \citet{Mazap02}.  
As noted in those papers, the spectrum is similar to that of SN~1997ef, 
with features that are broader than SN~1994I \citep{Iwamoto:94I}, 
but narrower than SN~1998bw.  
Prior to maximum light, the spectrum has few distinguishable features.
By maximum light, one can identify OI $\lambda$ 777.4, shallow Si II,
He I $\lambda$ 587.6 or Na I D and blue emission due to Fe II at
about 550 nm and 450 nm.  This evolution is reflected in the total
flux spectra shown here in Figs. 1 -- 3.  

NIR spectra show C I at 940, 1068 and 1069 nm over an extended period of time
(Gerardy et al. in preparation), but no He I at 2.058 $\mu$m on 8 March
\citep{Marion02IAU}. In addition, the NIR spectra show no
evidence for the strong Ni/Co blends at 1.6 to 1.8 $\mu$m that
are typical of SN~Ia.  As we will argue below, this suggests that
no radioactive elements have been lifted to or beyond the
photosphere at early times.  The early optical/NIR spectra are 
consistent with a carbon/oxygen atmosphere. 

\subsection{Interstellar Polarization}

To derive the intrinsic polarization due to the supernova atmosphere, we
first need to deduce the component due to interstellar dust. 
We use the method outlined in \citet{Wang:2001} that assumes that 
at each specific epoch, the polarization produced by the supernova ejecta 
has a single polarization position angle independent of wavelength. 
This method demands that the interstellar polarization falls along 
the dominant axis and must remain fixed independent of the epoch. 

We find the interstellar polarization to be at $Q\ =\ -0.26 \pm 0.05$, 
and $U\ = \ -0.55 \pm 0.05$.  We will adopt this as the best estimate 
of the interstellar polarization as shown by the black solid dot
in the upper panels of Figs. 1 -- 3.  In each of those panels,
the black solid line connects the chosen value of the ISP to the
origin of the Q-U plane.
The Galactic extinction is E(B-V) = 0.008 \citep{Schlegel:1998}, 
which would produce a maximum degree of polarization too small to account for
the 0.6\%\ interstellar polarization.  Most of the ISP and associated
extinction are thus most likely to be in the host galaxy.
This analysis yields a polarization position angle of 24\degree\ 
for the interstellar polarization.  It is not clear that this 
corresponds to any special direction in M74.

\subsection{Decomposition of the Polarization}

\citet{Wang:2001} outlined a method to decompose the observed polarimetry
into two components. On the Q - U plot, the two components correspond to
the polarized vectors projected onto the so-called dominant axis
(denoted by subscript ``d" below) and the axis orthogonal 
to the dominant axis (denoted by subscript ``o" below). 
The dominant axis can be defined from the aspherical distribution of 
the data points on the Q - U plane.  The dominant axis is derived
by a linear fit to the data points weighted by the observational
errors in the Q - U plane. The spectropolarimetry projected to the
dominant axis represents global geometric deviations from spherical 
symmetry whereas the vector orthogonal to the dominant axis 
represents deviations from the dominant axis. 
These components are defined as:
$$
P_d \ = \   (Q-Q_{ISP}) \cos \alpha\ + \ (U-U_{ISP}) \sin \alpha, \eqno(1)$$
and
$$P_o \ = - (Q-Q_{ISP}) \sin \alpha\ + \ (U-U_{ISP}) \cos \alpha, \eqno(2)
$$
where $P_d$ and $P_o$ are the polarization components parallel to
the dominant axis and orthogonal to that axis, respectively, and
\qisp\ and \uisp\ are the Stokes parameters of the interstellar
polarization.

In simple spheroidal models \citep{Hoeflich:1996,
WWH:1997}, the polarization is produced by electron
scattering through an aspherical atmosphere, and the spectropolarimetric
lines are formed  because of line scattering and depolarization. 
Such a geometry would produce a single line on the  Q - U plot, 
with the wavelength region that suffered the
most line scattering being most de-polarized and hence the closest to the
origin. The earliest data on SN~2002ap, from 3 Feb, approaches
such a situation.
 
%With the above choice of interstellar polarization, we can decompose the
%observed polarization into the dominant and orthogonal components. 
%The new coordinates are given by rotating the original coordinate systems 
%counter-clockwise so that the Q-axis overlaps the dominant axis in 
%the new coordinate system. The dominant axis is determined to be at 
%position angle -26.57\degree\ [CHANGE] through a linear fit
%to the data points on  the Q - U plot of 3 Feb. We have rotated the Q - U
%coordinate system by 26.57\degree\ [CHANGE] clockwise so that the 
%dominant axis points toward the center of the  data cluster on the 
%Q - U plot. 

We show in the top panels of Figs. 1 to 3 the observed data points 
in the Q - U plane. Each point represent a data pair corresponding to 
the Q - U vector at a different wavelength.  The wavelength of the data 
points in important intervals are encoded in color.
The data show remarkable evolution during the three epochs of observation.
The polarization also shows spectral features that can be identified
with features in the flux spectra. This firmly
establishes that SN 2002ap was intrinsically polarized
well before optical maximum.

The components parallel and  orthogonal to the dominant 
axis as given by $P_d$ and $P_o$ are plotted in the second
panels of Figs. 1 -- 3 where it can be seen that the $P_o$ component is
fairly flat except near the 750 nm feature.  Most of the 
continuum polarization is recorded in $P_d$. 
The continuum polarization (around 600 to 650 nm) in nearly zero
in the 3 Feb data, perhaps slightly positive by 7 Feb, and shows
a definite positive value, about 0.2 percent by 10 Feb.
This suggests that the global distortion was increasing with time.
The situation is more complicated, however, than can be 
interpreted with a simple spheroid model since other components
with different principle axes are revealed in the later data.

The third panels of Figs. 1 -- 3 show the total flux and the 
polarized flux in the dominant and orthogonal components.
Polarized spectral features are unambiguously detected around 
O II 777.4 nm, Na I D (or He I 587.6 nm) and in the blue around
400 nm where the feature is probably due to Fe II.
The wavelength at which the degree of polarization reaches maximum 
was around 730 nm on 3 Feb, moved to about 720 nm on 7 Feb, 
and to 700 nm on 10 Feb.

SN 2002ap exhibits features that are unlike those of
previously observed SN II and the subluminous SN Ia 1999by
\citep{Wang:1996, WWH:1997, Wang:2001, Leonard:2001a, Leonard:2001b,
Howell:2001}. On the Q - U plot, SN 1998S and SN 1999by showed 
well-defined linear features \citep{Wang:2001,Howell:2001} 
indicating relatively well-defined symmetry axes. 
SN~2002ap shows this behavior in the earliest data presented
in the top panel of Fig. 1 for the data of 3 Feb (-6 d).  
By 7 Feb (-2 d; top panel, Fig. 2), the dominant axis had shifted.
One interpretation is that the dominant axis has rotated by 
about 24\degree\ to become approximately parallel to the solid
line connecting the axis and the ISP point in the Q-U plane.
This would  indicate a change of orientation in the supernova 
ejecta of about 12\degree.  A rotation of axes of this sort could 
be due to the recession of the photosphere by different amounts
in different directions in response to the asymmetric velocity
field and density distribution \citep{HWW:1999}.
Alternatively, the locus of lower points in the first panel of Fig. 2
could represent a shift of the dominant axis in the first panel 
of Fig. 1 parallel to itself. This could occur if there is a 
growing wavelength-independent scattering that everywhere
increases the amplitude of the polarization of this dominant
component.  By 10 Feb (+1 d; top panel, Fig. 3) there has been 
a further shift of the dominant axis roughly parallel to that of 
3 Feb or 7 Feb and the development of a component nearly 
orthogonal to that dominant axis in the Q-U plane. 

\subsection{Polarization of the O I 777.4 Line}

The strongest feature in the polarization spectra is the line
at 740 nm.  This feature peaks in the dominant component
(second panel of Figs. 1 -- 3) at 1.1 percent, 
nearly 1.4 percent and 0.9 percent,
for 3 Feb, 7 Feb, and 10 Feb, respectively.  The feature
is smaller in the polarized flux plots (third panels), but still
prominent. This polarization feature narrows and its peak shifts
to the blue over the epoch observed.  \citet{Wang:ap02IAU} noted
that if this feature were the Ca II IR triplet, the implied velocity
would be about 45,000 \kms.  Closer study and especially the
time-dependent behavior reveals that Ca II is not the correct 
identification.  Rather, the behavior of this feature is most
consistent with the line being predominantly O I 777.4 nm that
is eaten away by the emergence of the Ca II IR triplet as 
the supernova evolves to maximum light.  With this identification,
the velocity is about 20,000 \kms.

The wavelength range that spans the O I feature also serves to define
the dominant component in the top panel of Fig. 1 as denoted
by the dashed line.  Note that there is some suggestion of two 
approximately parallel components with this dominant orientation, 
although the lower component (along Q $\sim$ -0.2) may not be 
statistically significant.  By 7 Feb, the shift or rotation
of the dominant component is basically a property of the O I line,
with little shift in other wavelengths.  Certainly, the original
component defined by the dashed line in Fig. 1 has disappeared
by the epoch represented in Fig. 2.  The top panel of Fig. 3 shows
again that the parallel displacement of the original dominant
component is once again primarily a manifestation of the
wavelengths corresponding to the O I line.  

The displacement of the O I component parallel to the original dominant
axis as the supernova evolves is presumably caused by the increase in 
the continuum polarization and perhaps by a shift in the geometry as 
the photosphere recedes by different amounts along different directions
\citep{HWW:1999}.  The line
is presumably polarized because of ``shadowing" of parts of the
asymmetric line emitting region by portions of the ejecta that
remain optically thick.  Quantitative analysis of such effects
will require detailed radiative transfer models that we postpone
to later work.
 
By 10 Feb, shortly after maximum light, there is obviously a
new absorption component at 770 nm as shown in the middle and bottom
panels of Fig 3.  We interpret this as due to the Ca II IR triplet, 
in which case it also represents a velocity of about 20,000 \kms. 

The interpretation of the double dips at about 680 nm and 770 nm
in the orthogonal component (second panels) in Figs. 1 and 2 is
not clear.  Their amplitude depends rather sensitively on
the choice of the ISP.

\subsection{Polarization of Na I D}

We believe the feature at 550 nm is most likely to be 
Na I D.   This feature becomes especially
prominent on 7 Feb in the polarized flux (panel 3, Fig. 2),
but then less so by 10 Feb (panel 3, Fig. 3).  
The wavelengths corresponding to this feature cluster
around Q $\sim$ 0 and U $\sim$ 0 in the top panel of Fig. 1.
%This is presumably because with its widely spaced levels,
%helium is less effective at depolarizing the scattering continuum
%flux than O or Fe.  
By 7 Feb and 10 Feb, the corresponding wavelengths are still
tightly bunched in the Q-U plane, but the bunch has shifted slightly 
along the direction of the solid line in Figs. 2 and 3.   

This feature could be either He I 587.6 nm or Na I D.  We favor
the latter because of the lack of other evidence for helium.
In particular, the He 2.08 $\mu$m line was not observed in NIR spectra
taken several weeks later, 8 March \citep{Marion02IAU}. 
This may be because the abundance is low and the helium is confined 
to the outermost layers, but there cannot be a strong exposure of any 
helium to radioactive excitation at that time \citep{Swartz87M}.
Other constraints on helium arise from earlier NIR spectra, some 
contemporaneous with the optical spectropolarimetry we present here, 
that also fail to show any evidence for the He 2.08 $\mu$m line
(Gerardy et al. in preparation).  If the line at 550 nm were
He I 587.6, then the associated line at 2.08 $\mu$m should
be unambiguous. We note that a similar feature at 550 nm was seen in
SN~1993J. \citet{Hoeflich:1996} argue that an NLTE enhancement of
a factor of $10^9$ would have been required to account for the feature
as He I 587.6, making that feature more likely to be Na I D as well.
The Na is surely out of NLTE, and quantitative models are
required to explore whether Na can account for the behavior seen
here in SN~2002ap, or in SN~1993J .       

\subsection{Polarization of Fe II}

The third prominent feature in the polarization spectra is
that at about 450 nm. In Fig. 1, this feature is likely to 
be Fe II.  A similar feature was seen in the spectropolarimetry
of SN~1993J \citep{Trammell:1993, Hoeflich:1996}.  This Fe is likely to
just be the primordial iron present within the outer carbon/oxygen
envelope since features associated with iron freshly synthesized 
by radioactive decay are not seen (see \S 4 below).  This polarization
peak has essentially vanished in the later data.
In Fig. 1, this feature seems to have a spread parallel to
the dominant axis in the Q-U plane, displaced from the location of the 
sodium feature and perhaps coincident with the lower Q component
of the O I line, inasmuch as that data marginally resolves into
two parallel components (\S 3.3).  
By 7 Feb, this displacement has disappeared 
and any remaining iron feature is essentially contiguous with the sodium
feature at Q $\sim$ -0.2 and U $\sim$ -0.2.  

Although the signal to noise is degraded, there is a dramatic change 
in this blue feature by 10 Feb when it seems to define a new axis 
with a very different orientation from the dominant axes of 3 Feb 
and 7 Feb.  This new feature may be associated with Ca II H \& K,
as will be discussed in the next section.

We note that this polarization feature at 400 nm on 3 Feb
is interrupted by a sharp minimum at 390 nm in the polarization 
spectra of Fig. 1.  This narrow feature might be interstellar 
Ca II H \& K in the Galaxy or in M74 or in the circumstellar matter 
near the supernova.  There is no easily discernible feature in the 
total flux spectrum.  In addition, the feature appears to be transient.  
While the data do not extend sufficiently far to the blue to define 
this region on 7 Feb, there is little sign of this feature in the 
polarization data of 10 Feb.  This might suggest that this feature 
is connected with the circumstellar matter very near the supernova, 
but it may be that the feature is an artifact of the sensitivity near 
the blue end of the spectrum.

\section{The Geometric Structure of SN 2002ap}

The polarization evolution allows us to extract some useful information 
about the ejecta structure.  As a first approximation, we will assume 
that close to the photosphere the ejecta can be modeled as a prolate or 
oblate spheroid.  We define the asphericity of the spheroid as the ratio 
of the major to minor axes ratio minus 1, expressed as a percentage.  
The relation between this geometric asymmetry and the angular distribution 
of the luminosity for an electron-scattering atmosphere is taken
from \citet{Hoeflich:1991} (see also \citealt{WWH:1997}). 
This spheroid approximation may
reproduce the orientation and distortion along the dominant axis of asymmetry.  
The photosphere itself is likely to be more irregularly structured, so the 
spheroid is only an approximation.  In particular, in an asymmetric 
explosion, during homologous expansion the location of the photosphere
will depend on the profile of density as a function of velocity and
that profile will depend on angle.  The photosphere will recede at
different rates in different directions yielding isodensity
contours that are not simply ellipsoidal.  An illustration
is given in Fig. 4 (from H\"oflich, Wang \& Wheeler 1999)

The maximum degree of continuum polarization is nearly zero on
3 Feb implying little distortion of the photosphere. This
suggests that the outer envelope has not been penetrated by
any bipolar flow or other perturbation that would give a severe distortion 
and a large polarization, but the low polarization could also be 
consistent with viewing down the symmetry axis.  
Since the continuum polarization evolves
to higher values, the latter cannot be the full explanation of the early
low polarization, although orientation effects clearly play a role.

By 10 Feb the continuum polarization has increased to about 0.2 percent.  
This would be consistent with an asphericity of $\sim$ 10 - 15  percent 
for an ellipse observed along the equator \citep{Hoeflich:1991, Howell:2001}. 
This polarization amplitude must be interpreted carefully, since it is also
accompanied with strong evolution of the polarization spectra
and with the rotation or displacement of the dominant axis and the
appearance of new axes in the Q-U plane.

How, then, might one account for a principle polarization
axis of SN 2002ap seen on 3 Feb, the possible rotation by 24\degree\ on
7 Feb, and the appearance of a new axis of orientation on 10 Feb?
Detailed modeling with 3-D radiative transfer is required to
fully understand the polarization data \citep{Hoeflich:1996, WWH:1997,
Howell:2001}. This analysis will be presented elsewhere. Here
we will discuss some of the physical possibilities.  

Important constraints come from the combined consideration of the
polarization data and the NIR spectroscopy (Gerardy et al. in preparation).
We have identified here, with the critical help of the polarization
evolution, the feature due to O I 777.4 nm.  The lack of evidence
for the He 2.08 $\mu$m line argues that the feature at 590 nm
is Na I D.  The NIR spectra do reveal strong lines of C I at 940.5, 1068.3
and 1069.1 nm.  On the other hand, the 
NIR spectra fail to show the strong lines blends of nickel and cobalt 
at 1.6 - 1.8 $\mu$m that are routinely shown by Type Ia supernovae 
(\citealt{WheelerIR:1998, Hoeflich99byIR}; Marion et al. in preparation) 
that represent freshly synthesized radioactive decay products.
This combined information suggests that the outer portions of the
atmosphere of SN~2002ap are rich in carbon and oxygen with a small, if any, 
layer of helium, as expected for an SN~Ic.  The lack of
the nickel/cobalt features at 1.6 - 1.8 $\mu$m strongly suggests
that radioactive matter has not penetrated the photosphere at the
earliest epochs reported here.  Thus we conclude that there has
been a major perturbation of the ejecta in a manner that leads to
a dominant axis of symmetry as revealed in the Q -U plane of the
early data (3 Feb), but that the predominantly bi-polar flow
that shaped such an asymmetry has not penetrated the surface of
the star during the explosion.

The photosphere can potentially evolve in a rather complex way
in response to density and velocity anisotropies in the ejecta,
as illustrated by \citet{HWW:1999}.  In that example, high
velocities along the axis of symmetry led to especially rapid
recession of the photosphere along the axis, leading to a roughly
quadrupolar isodensity contour structure at intermediate times
(Fig. 4). This evolution in the isodensity
contours of a nominally oblate shape could lead to the shift in 
the orientation of the dominant axis and could help to explain
some of the behavior seen in Figs. 1 -- 3.
  
The emergence of a second dominant axis in the 10 Feb data clearly
reveals a separate kinematic component.  Some of this 
component may be reflected in the O I and Ca II features, but
it is predominantly a feature in the blue. This could be the
emergence of Ca II H \& K in the blue at the same time that
the effect of the Ca II IR triplet is being manifest in the red.
In jet-induced models, the calcium is expected to be confined to
equatorial tori \citep{KhokhlovKor, HoeflichTex, Wang87A:2002}. 
This new axis
is rotated compared to the dominant axis of the 3 Feb data by
about 100\degree\ in the Q-U plane, corresponding to about
50\degree\ in physical space.  This is consistent with having 
an outer photosphere with a ``quadrupolar" structure reflected in
the oxygen (and presumably carbon) that is tilted by about
50\degree\ with respect to the equator and then a deeper torus
of calcium-rich matter on the equator. 

\section{Discussion and Conclusions}

Understanding the 3-D nature of core collapse supernovae, especially
``naked core" events is an important key to the underlying physics, 
and polarimetry is the best tool to probe multi-dimensional effects.

SN~2002ap resembles other Type Ib/c events, but displays rather
broad lines near maximum.  The distinct polarization spectra show
that the ejecta are asymmetric.  The premaximum polarization spectra
reveal a strong feature of O I 777.4 nm and probably Na I D,
both moving at about 20,000 \kms.  Near maximum light, a new dominant
axis is revealed as defined, most likely, by calcium again with
a velocity of about 20,000 \kms.   These observations can be
accounted for in a model in which a bi-polar flow from the exploding
core imposes a predominantly prolate velocity structure and oblate
density structure.  The outer photosphere at early times can
thus have a complex structure with an effective dominant axis that
is intermediate between the axis of symmetry and the equator.  Later
observations would reveal a more predominantly oblate structure
oriented along the equatorial plane, as observed.  This picture is
supported by the lack of observations in the NIR of freshly
synthesized nickel and cobalt as seen in Type Ia, suggesting that
the bi-polar flow has induced velocity and density asymmetries, but
has not penetrated the photosphere at early times so that 
these freshly synthesized elements are not observed.  

We find no evidence for substantial amounts of matter moving at
45,000 \kms, which was mentioned as a possibility by \citet{Wang:ap02IAU}.
The relatively high velocities that are observed,
about 20,000 \kms, broaden the lines compared to some other 
Type Ic, for instance SN~1994I \citep{Iwamoto:94I}.  This can be accounted for
by a combination of different ejecta mass and an asymmetric 
ejecta. In general, a higher ejecta mass and perhaps a steeper
density gradient at the photosphere, can prolong the phase
in which the photosphere resides in higher velocity material.
We note also that the earliest spectra of SN~2002ap were obtained 
substantially before maximum light, a rare happenstance for
SN~Ic events.  In addition to this basic effect of the
ejecta mass, the asymmetry of the ejecta will bring line-of-sight
effects.  As shown by \citet{HWW:1999} the 
isodensity contours can be rather irregular, with a quadrupole-like
distortion as the ejecta expand more rapidly along the poles,
even for an ellipsoidal density distribution.  For a more realistic
bi-polar explosion the density contours can be even more
complicated.  Thus there may be special viewing angles along which
the homologous expansion is relatively slow, leading to a delayed
recession of the photosphere and hence to a larger line-of-sight
photospheric velocity at a given epoch.  In any case, it is
clear that while a proper interpretation of polarized, asymmetric ejecta
is tricky, it is certainly risky to interpret observations of
asymmetric ejecta in terms of
a spherically symmetric model and to take the results literally.
 
The luminosity of SN~2002ap was consistent with a normal
Type Ib/c, there was no \grb, and the radio
observations are consistent with only a modest ejection of
high-velocity matter \citep{Berger:2002}.  
The relatively high early photospheric velocities of SN~2002ap
with respect to other SN~Ic, may then just be a question of
epoch of observation, mass of ejecta, and orientation angle of the 
observer, and not a hint of exceptional properties.

If one can interpret SN~2002ap as a variation on the ``normal"
theme of Type Ib/c supernovae, then perhaps one should reconsider
the general category of ``hypernovae."  In the context of supernovae
(as opposed to Paczy\'nski's original definition), SN~19998bw is the
prototype of this proposed new category with large kinetic
energy, ejecta mass, nickel mass and luminosity attributed to
it in the context of spherically-symmetric models.  \citet{Iwamoto98bw}
determined that the exploding carbon/oxygen core
had a mass of 12 to 15 \Msun\ with kinetic energy in the range
20 to 50$\times10^{51}$ erg and mass of \ni\ in the range
0.6 to 0.8 \Msun. \citet{Woosley98bw} favored a carbon/oxygen
core of 6 \Msun, a kinetic energy of $22\times10^{51}$ erg
and a mass of \ni\ of 0.5 \Msun.  Both of these models had
some problems reproducing the spectra or light curves.  
This class of models also had problems reproducing the late
time tails.  The predicted light curves tended to drop too
quickly.  Another intrinsic problem is that these spherical
models with very high kinetic energy imposed at the base of
the ejecta predict very high minimum velocities.  The latter
two problems are related.  The rapid expansion of the innermost
layers cause them to thin out, thus decreasing the \gr\ deposition
and causing the drop in the light curve.  

By contrast, a somewhat schematic model by \citet{HWW:1999}
showed that a good fit could be obtained to the multi-color light curves
of SN~1998bw near maximum with an asymmetric model.  This model required
somewhat higher kinetic energy and \ni\ mass, $2\times10^{51}$ erg
and 0.2 \Msun, respectively, but not sufficiently large values to 
warrant defining a new category of ``hypernovae." Furthermore, models of 
jet-induced supernovae \citet{Khokhlov:1999} showed that
asymmetric models would undergo circulation and inflow that
would yield much smaller minimum velocities than the corresponding
spherical models.  These small minimum velocities were revealed
by late time observations of SN~1998bw (Patat et al. 2001).
The slow moving matter will, in principle, allow for greater
trapping of \grs\ at later times and hence flatter late-time
light curves requiring less \ni, all else being equal.   
\citet{Maeda:2002} also conclude that asymmetric explosions can 
lead to slower moving matter near the center, but they only explored
very high mass carbon/oxygen cores in the context of SN~1998bw, 
and hence concluded once again that high kinetic energy is required 
to match SN~1998bw.  

Estimates of the amount of \ni\ required to power the
light curve of SN~1998bw have also evolved with time.  Based
on the first 123 days, \citet{McKenzie:1999} estimated
a minimum of 0.2 \Msun.  While this number agreed with the
estimate of \citet{HWW:1999}, it was considerably less than
the estimates based on the first spherically-symmetric light
curve models.  A later, very careful analysis of the late-time
light curve, especially HST photometry at 800 to 1000 d, by
\citet{Sollerman:2002} showed that a simple model could
fit the light curve with only 0.3 \Msun\ of \ni.  More
sophisticated models allowing for the slowly expanding and
efficiently trapping inner layers expected from asymmetric
models might very well do the job with less \ni.  There is,
in any case, little rationale for the early large estimates
of \ni\ mass based on spherical models.  In this context, we 
note that the ejected \ni\ mass for normal SN~II is reported 
to range as high as 0.3 \Msun\ \citep{Schmidt:1994}. \citet{Germany:2000}
report an implied \ni\ mass of 2.6 \Msun\ for SN~1997cy, but this
number is so discrepant with other determinations of the
\ni\ mass that we are tempted to assume a substantial proportion
of the luminosity of this event arise from circumstellar interaction.
More events like SN~1997cy will have to be discovered to
clarify this situation. 

If one removes the
seeming excessively large nickel mass and can, in principle,
account for the observed optical luminosity of SN~1998bw by
invoking the angle-dependent luminosity expected from 
asymmetric models, and can account for the large photospheric
velocities by a judicious choice of ejecta mass and viewing
angle, then one still has the active possibility that
SN~1998bw was an event at the high end of the ``normal" range
of SN~Ic supernovae with twice the kinetic energy and
twice the nickel mass of ``normal" SN~Ic.  One does
not necessarily need to define a whole new class of ``hypernovae"
to account for it.  Similar arguments can account for events
that are perceived to be intermediate between SN~1998bw and
SN~1994I, such as SN1998ef.

Even if one does not accept that one can account for the
observations of SN~1998bw, SN~1998ef, and SN~2002ap in
terms of mild variations and rather extreme asymmetries, 
the polarimetry shows that SN~1998bw and SN~2002ap, at least,
were asymmetric and this must be taken into account.  
The \grb\ community recognized the key role that collimation plays in 
\grbs\ by adopting the language of ``isotropic equivalent energy"
in appropriate contexts when spherically symmetric models were
applied to asymmetric situations.  Perhaps it is time for
the supernova community to do that as well and learn to 
speak of isotropic equivalent energies, ejecta masses, and
nickel masses when spherically-symmetric models are 
inappropriately applied to asymmetric situations.

\acknowledgements The authors are grateful to the European
Southern Observatory for the generous allocation of observing time. 
They especially thank the staff of the Paranal Observatory for 
their competent and untiring support of this project in service 
mode.  This work was supported in part by NASA Grant NAG5-7937 to PAH
and NSF Grant AST 0098644 to JCW.

%\bibdata{REF.bib}
%\bibliography{REF}

\newpage

\figcaption[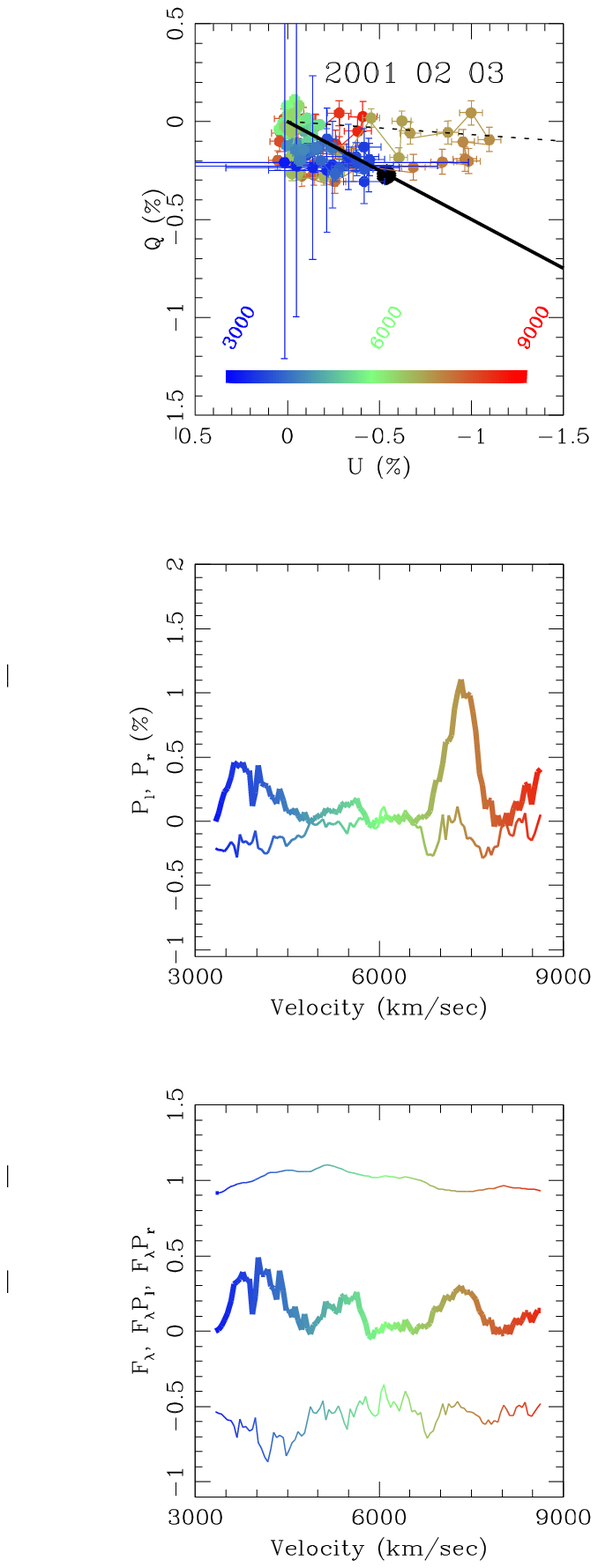]{
Spectropolarimetry of SN 2002ap on 2002 3 Feb, 6 days before V maximum. The
Stokes parameters are rebinned into 15 \AA\ bins. An interstellar
polarization component is subtracted from the observed Stokes 
Parameters so that the data points represent intrinsic polarization due 
to the supernova. The assumed interstellar polarization is shown as 
the solid dot in the Q - U plot (top panel).  Without subtraction
of the interstellar component, the origin of the coordinates would be
centered at this solid dot. The solid line represents the axis from
the origin through the value of the ISP.  The dashed line illustrates 
the dominant axis shifted to the origin of the Q - U plot.  
The polarization spectra (middle panel) and polarized flux (bottom
panel) show conspicuously polarized spectral features corresponding to 
Fe II, Na I D, and O I 777.4 nm (see text).  The wavelength color code 
is presented at the bottom of the top panel.
\label{Feb3}}

\figcaption[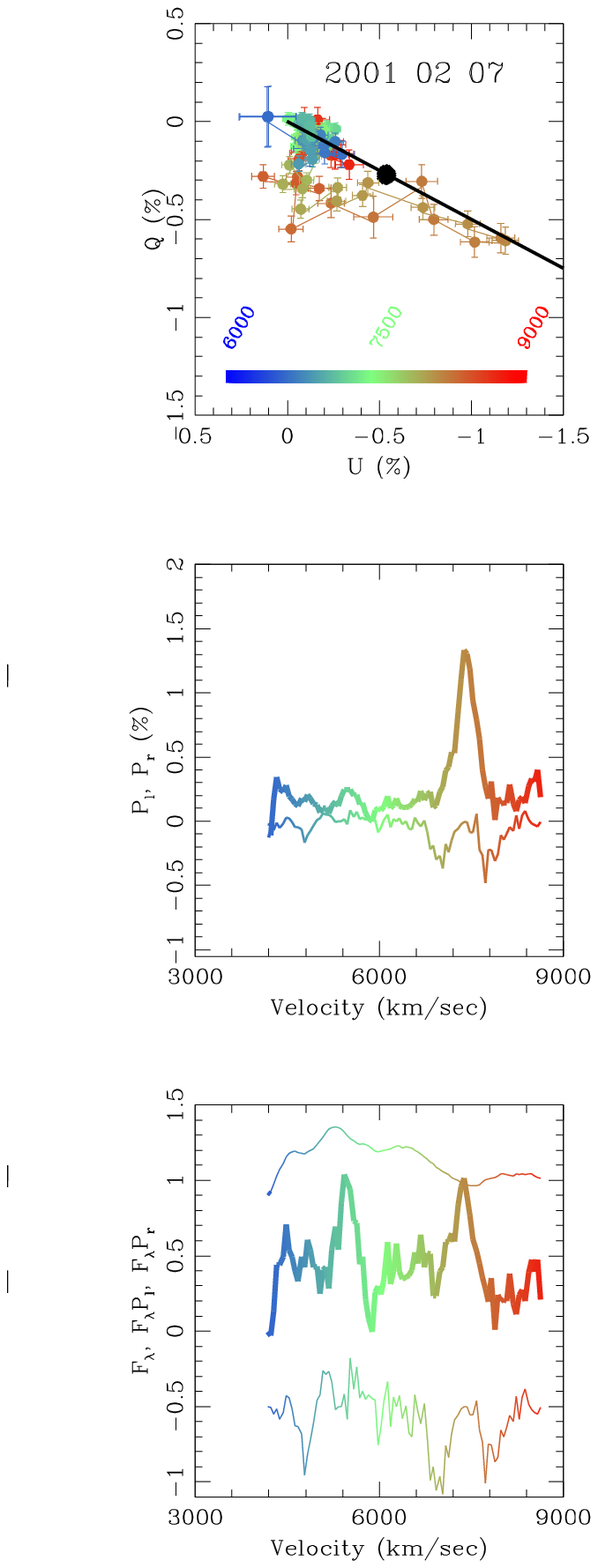]{
Spectropolarimetry of SN 2002ap on 2002 7 Feb, 2 days before V maximum
(see Fig. 1).  There has been a distinct shift and perhaps a rotation
of the dominant axis in the Q-U plane (top panel).  The feature
corresponding to O I 777.4 nm is the only significant feature in
the polarization spectrum (middle panel), although when weighted by
photon flux, the polarized flux spectrum (bottom panel) still shows
evidence of the Na I D feature.
The wavelength color code is presented at the bottom of the top panel.
\label{Feb7}}

\figcaption[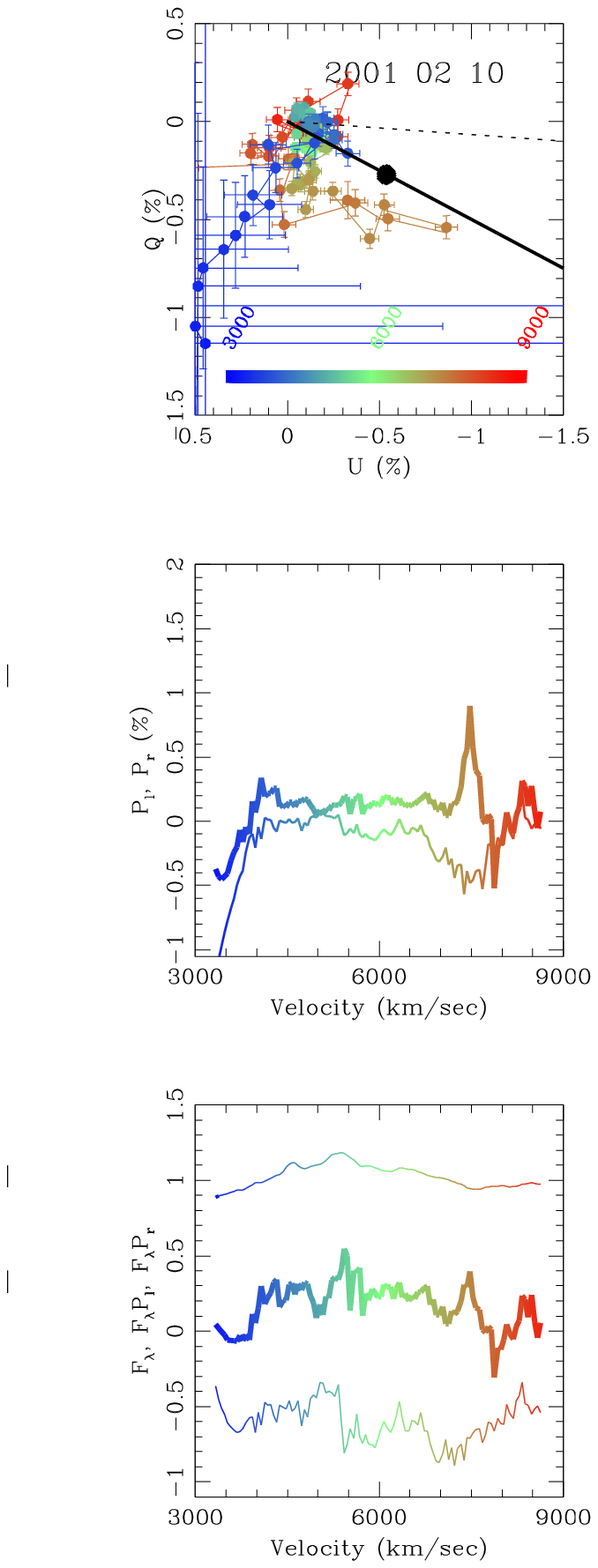]{
Spectropolarimetry of SN 2002ap on 2002 10 Feb, 1 day after V maximum
(see Fig. 1). The original dominant axis has shifted further in the
Q-U plane and a new axis, probably defined by Ca II H \& K and the Ca II
IR triplet appears at an angle of approximately 110\degree\ with
respect to the original dominant axis (dashed line).  The peak
due to O I 777.4 nm has shifted to the red as the Ca IR triplet
has begun to cause significant depolarization to the blue of 
the O I feature.
The wavelength color code is presented at the bottom of the top panel.
\label{Feb10}}

\figcaption[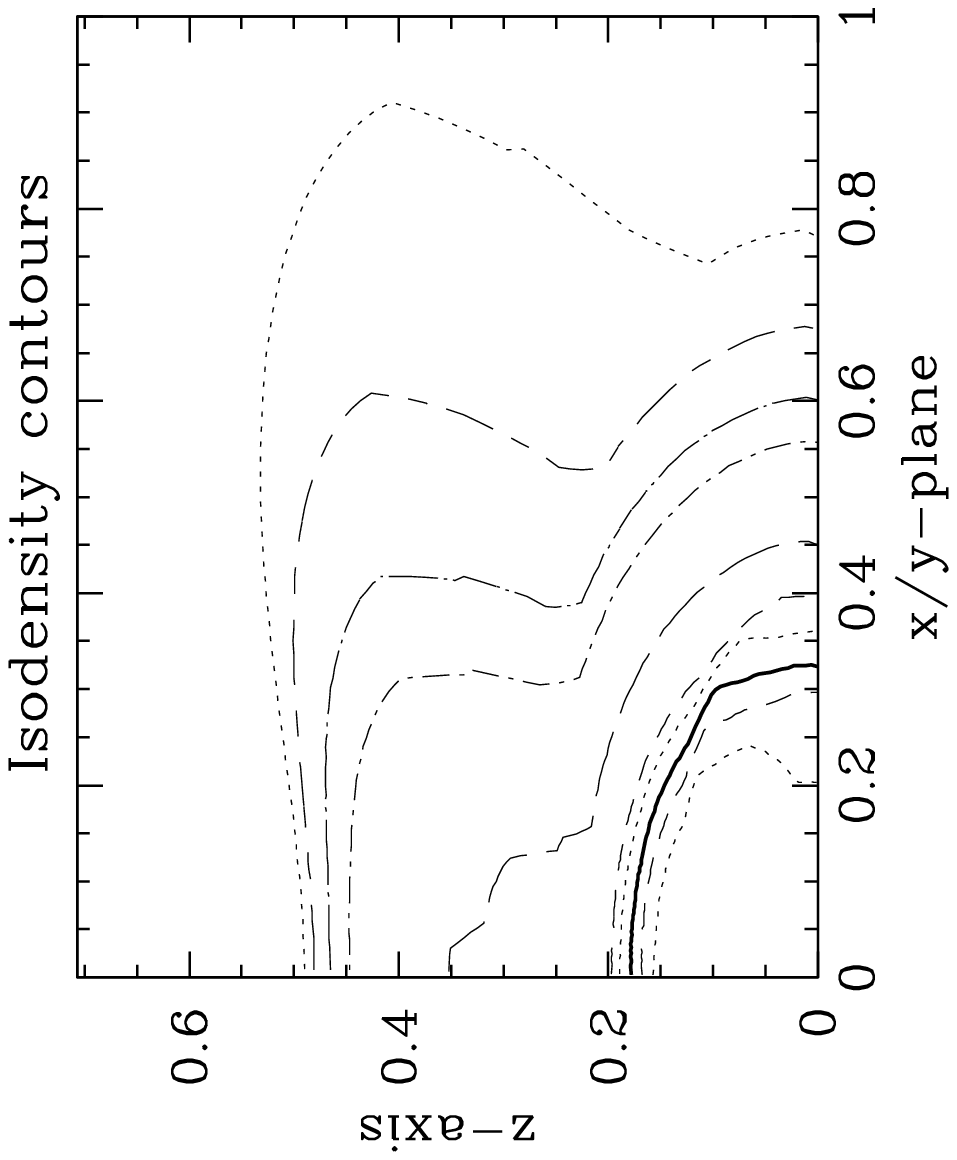]{
Isodensity contours for a model with oblate geometry,
spherical isovelocity contours, and prolate Lagrangian mass
elements.  The contours are defined
by the location in the equatorial plane of mass elements that
correspond to mass fractions in a corresponding spherical model
of 0.1, 0.3, 0.5 (thick line), 0.6, 0.7, 0.8, 0.9, 0.93 and 0.98. 
The recession of the photosphere by different amount in different
times coupled with special viewing angles can lead to blocking
of parts of the photosphere by optically thick portion and
hence to apparent rotation of the dominant axis even though
the symmetry axis remains fixed. (From H\"oflich,
Wang \& Wheeler 1999).
\label{contours}}

\newpage

\plotone{fig1.ps}

\newpage

\plotone{fig2.ps}

\newpage

\plotone{fig3.ps}

\newpage

\plotone{fig4.ps}

\end{document}